\documentclass[showpacs,aps,prl,twocolumn,superscriptaddress,floatfix]{revtex4}
\usepackage[]{braket}
\usepackage[]{graphicx}
\usepackage{epsf}
\usepackage{subfigure}
\usepackage{rotating}

\newcommand{\op}[1]{\hat{#1}}

\newcommand{\hdot}[0]{\op{H}_{dot}}
\newcommand{\hwire}[0]{\op{H}_{wire}}
\newcommand{\hint}[0]{\op{H}_{int}}

\newcommand{\imaginaryi}[0]{\mathrm{i}}

\newcommand{\mcom}{\mbox{\ ,}}

\begin{document}

\title{Readout of solid-state charge qubits using a single-electron pump}

\author{C. Hines}
\affiliation{School of Physics, The University of Western Australia, Perth 6907, 
Australia}
\author{K. Jacobs}
\affiliation{Centre for Quantum Computer Technology, Centre for Quantum Dynamics, 
School of Science, Griffith University, Nathan 4111, Australia}
\affiliation{Department of Physics, University of Massachusetts at Boston, 100 Morrissey Blvd, Boston, MA 02125, USA}
\author{J. B. Wang}
\affiliation{School of Physics, The University of Western Australia, Perth 6907, 
Australia}


\begin{abstract}  
A major difficulty in realizing a solid-state quantum
computer is the reliable measurement of the states of the quantum registers. In
this paper, we propose an efficient readout scheme making use of the resonant
tunneling of a ballistic electron produced by a single electron pump. We treat
the measurement interaction in detail by modeling the full spatial
configuration, and show that for pumped electrons with suitably chosen energy
the transmission coefficient is very sensitive to the qubit state. We 
further show that by using a short sequence of pumping events, coupled with a 
simple feedback control procedure, the qubit can be measured with high
accuracy. 
\end{abstract}


\maketitle


Reliable measurement of qubits is a key issue in quantum computation, either
in the output register~\cite{DiVincenzo98} or as a central ingredient in the computation~\cite{Raussendorf01}.  
For solid-state qubits, such as quantum dots~\cite{Loss98} or Cooper-pair boxes (CPBs)~\cite{Makhlin01}, a number of measurement techniques have been proposed. One approach is to place the qubit adjacent to a single electron transistor (SET), superconducting SET (SSET), or quantum point contact (QPC)~\cite{Korotkov, Makhlin00, elattari2000, aassime2001, wiseman2001, goan2001, jefferson2002, Clerk02, kinnunen2003, Emiroglua03, buehler2003, Averin05, Clerk06}). In this method the state of the qubit affects the tunneling current through the SET or QPC, and this current provides the read-out. Much experimental work on this method has been performed by the group of Clark~\cite{Buehler05,Buehler06}, and the measurement of single qubits with an SET has been realized by the groups of Nakamura~\cite{Yamamoto03,Pashkin03} and Williams~\cite{Gorman05}. Such a measurement has also been realized by Hayashi {\em et al.} by directly inducing tunneling from the double dot~\cite{Hayashi03}. However, none of these has yet been realized as a ``single-shot'' measurement. Two further methods for measuring a charge qubit have been both proposed and realized as single-shot measurements. The first is a scheme by Vion {\em et al.}~\cite{Vion02} in which the state of a CPB is converted to a supercurrent. The second has been implemented by the group of Martinis {\em et al.}~\cite{McDermott05,Katz06}, and is a partially destructive measurement in which a tunneling event is induced by a transition to a third level. A further measurement scheme using a stripline resonator has also been suggested by Sarovar {\em et al.} \cite{Sarovar05}. 

Here we consider a read-out scheme for a qubit that employs a single-electron 
pump~\cite{Likharev2001, jehl2003}. In this case, the measurement is obtained by passing a controlled
sequence of single electron pulses through a quantum wire placed adjacent to
the qubit. We take the qubit to be formed by a double quantum-dot, and in this 
case we can treat the interaction of the electrons with the system in detail, allowing 
us to characterize the disturbance to the system. 
We find that this disturbance does significantly limit the amount of
information which can be extracted with the raw measurement. However, because
the pump allows us to control the sequence of electron pulses, the qubit may
be manipulated by the application of unitary gates between the pulses. We show
that, so long as the pumped electrons have a reasonably well defined momentum,
such gates may be applied in a process of feedback so as to correct most of the
unwanted disturbance. This provides a near-perfect von Neumann measurement of
the qubit within a small number of pulses. We note that sequences of ``pulsed'' measurements, and sequential correction using feedback has also been considered by  Jordan {\em et al.} for measurements with a QPC~\cite{Jordan06b,Jordan06}.  

Our scheme is similar to those involving an (S)SET or quantum point contact, in that the information is extracted as electrons pass by the qubit. However, analyses of these schemes have invariably been performed assuming a simple interaction between the qubit and the probe system. This interaction is assumed to be proportional to the product of a Pauli operator for the qubit and an operator for the probe system~\cite{Korotkov, aassime2001, buehler2003, jefferson2002, kinnunen2003, Makhlin00, elattari2000, wiseman2001,goan2001, Clerk02,Clerk06}. The accuracy of this approximation will likely depend on system parameters (such as the well-depth of the quantum dot), and any deviation from this form will in general cause the passing electrons to disturb the system. Since many electrons are required to read out the qubit, a small unwanted disturbance by each electron could potentially impose a significant limit on the measurement. An advantage of our scheme is that the electrons are ballistic, and this allows us to model the Coulomb interaction between the probe electrons and the electron in the quantum dot in detail.

\begin{figure}[t]
\centering
\includegraphics[width=3in]{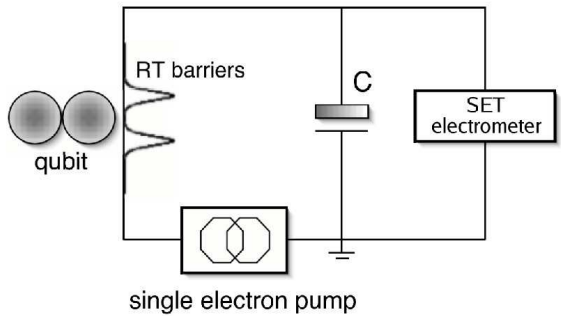}
\caption{A schematic diagram of the readout system. The electron is pumped 
through the RT-barriers to the capacitor C, the charge on which 
is measured by an SET electrometer.}
\label{fig:schematic}
\end{figure}

The readout configuration consists of a nano-wire connected to a single
electron pump~\cite{Likharev2001, jehl2003} and resonance tunneling (RT) barriers placed near
the qubit, which consists of a single electron in two coupled double quantum
dots separated by a central barrier. Localization of the electron on the right
dot (i.e. closer to the wire) represents the state $\ket{1}$, while localization on 
the left dot represents the state $\ket{0}$. Figure~\ref{fig:schematic} shows a 
schematic representation of the configuration. An electron is pumped through the 
RT barriers to the capacitor C. The charge on C is then measured using an
SET electrometer to determine whether the electron was reflected or transmitted
by the barriers, and this provides information regarding the charge state of the
dot. Since the electron on the capacitor is well separated from the rest of the
system, and since its precise state is unimportant, the measurement is robust
against any details of the interaction with the SET electrometer, which can
perform the charge measurement with very high accuracy~\cite{martinis94}. The
single electron pump produces electrons with a reasonably well defined and
tunable energy, and the device is operated in the ballistic transport regime. 

In modeling the system we include in the Hamiltonian the kinetic energy
operators of the electron in the nano wire and the electron in the quantum
dots, the Coulomb interaction between the two electrons, the potential of the
RT barriers for the electron in the nano wire, and the confinement potential
for the electron in the quantum dots.  Apart from adopting the effective mass 
and permittivity approximation to account for lattice and screening effects due 
to the solid enviroment and considering the spin-related relativistic terms negligibly 
small, this model includes all perceptible interactions in the system and allows all 
possible states for the qubit electron in the coupled quantum dots. It is therefore
expected to reflect accurately the quantum dynamics of the system during the
pumping of the electron across the RT barriers. The measurement of the charge
on the electrode at the completion of a pump cycle can be performed with great
accuracy~\cite{martinis94}, and may therefore be described as a von Neumann
measurement which distinguishes the sign of the final momentum of the pumped
electron. 

{\em Details of the Model:} The joint system composed of the double quantum dot
and the electron in the nano-wire is described by the wavefunction $\psi ({\bf
r}_1,{\bf r}_2,t)$, where ${\bf r}_1 = (x_1,y_1)$ is the position of the
electron in the coupled quantum dots, ${\bf r}_2 = (x_2,y_2)$ is the position
of the electron in the nano wire, and $t$ is time. We will often drop these
arguments in the following in order to keep the notation compact. The
Schr\"{o}dinger equation that governs the time-evolution of the system
wavefunction is 
\begin{equation}
\mathrm{i}  \hbar \frac{\partial \psi }{\partial t} = \mathcal{H} \psi , 
\end{equation}
where the system Hamiltonian is $\mathcal{H} =  \hdot +
\hwire + \hint$, $\hdot =  -\hbar^2 / (2m^*) \nabla_{{\bf r}_1}^2 + V_{dot}({\bf r}_1)$  is the Hamiltonian for the electron in the coupled dots, 
$\hwire = -  \hbar^2 /(2m^*) \nabla_{{\bf r}_2}^2 + V_{wire}({\bf r}_2)$ is the Hamiltonian for the electron in the wire, and 
$\hint = 1/ (4\pi\epsilon |{\bf r}_1 - {\bf r}_2|)$ describes the
interaction between the two electrons. The effective mass of the mobile
electrons is denoted by $m^*$, and $\epsilon$ is the effective permittivity.
If, for example,  the qubit system and the nano wire detector are
built in an AlGaAs/GaAs interface, the effective mass and permittivity are
given by $m^*=0.0667 m_e$ and $\epsilon=12.9\epsilon_0$~\cite{tanner1995}.   

To simulate the evolution of the system we use the Chebyshev-Fourier scheme as
detailed in~\cite{wang1999}. Briefly, this method approximates the exponential
time propagator by a Chebyshev polynomial expansion
\begin{equation}
  \label{eq:Chebyshev}
  \psi(t+\Delta t) = e^{-\imaginaryi(\mathcal{E}_{u}+\mathcal{E}_{l})\Delta t/2} 
				     \sum_{i=0}^{\mathcal{N}}a_{i}(\alpha) 
				     T_{i}(-i\tilde{\mathcal{H}})\psi(t) \mcom
\end{equation}
where $\mathcal{E}_{u}$ and $\mathcal{E}_{l}$ are the upper and lower
bounds on the energies sampled by the wavepacket, 
$\alpha=(\mathcal{E}_{u}-\mathcal{E}_{l}) \Delta t /2$,
$a_{i}(\alpha)=2J_{i}(\alpha)$ except for $a_{0}(\alpha)=J_{0}(\alpha)$,
$J_{i}(\alpha)$ are the Bessel functions of the first kind, and $T_{i}$ are the
Chebyshev polynomials. To ensure convergence, the Hamiltonian must be
normalized according to $\tilde{\mathcal{H}}=[2\mathcal{H}-
\mathcal{E}_{u}-\mathcal{E}_{l}]/(\mathcal{E}_{u}-\mathcal{E}_{l})$.

\begin{figure}[t]
\centering
\includegraphics[width=2.6in]{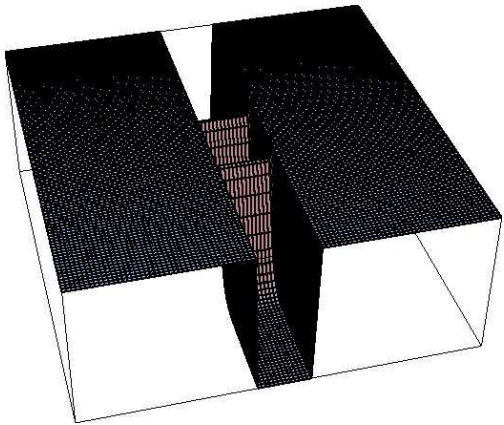}
\caption{Nanowire with the RT-barriers as defined by Eq (4).}
\label{fig:wire}
\end{figure}

We choose the computational basis states of the qubit, $\ket{0}$ and
$\ket{1}$, to be mutually orthonormal linear combinations of the lowest two
energy eigenstates of the double dot system. Since these
states are not the energy eigenstates, the system will oscillate between them,
and we must therefore ensure that the duration of the measurement is very small
compared to this oscillation time.
We model the double dot using the potential 
\begin{widetext}
\begin{eqnarray} 
  V_{dot} (x_1,y_1) & = & -V_0 \; exp \left( -\frac{m^*}{2 V_0} \left( x_1^2+(y_1-y_c)^2 \right) \omega^2 \right) -V_0 \; exp \left( -\frac{m^*}{2 V_0} \left( x_1^2+(y_1+y_c)^2 \right) \omega^2 \right) \!\! , 
\end{eqnarray} 
where $y_c=143~\mbox{nm}$, $V_0=5.99~\mbox{meV}$, and $\hbar
\omega=0.818~\mbox{meV}$ describe a typical double-dot.   
The wire RT barriers, as shown in Figure~\ref{fig:wire}, are described by 
\begin{eqnarray}
  V_{wire}(x_2,y_2) & = &  \frac{v_x}{\cosh^2((x_2-r)/s)} + \frac{v_x}{\cosh^2((x_2+r)/s)}  + v_y \left( 1 - \Theta(y_2+d+\delta y_2) \Theta(-y_2-d+\delta y_2) \right) \mcom
\end{eqnarray}
\end{widetext}
where $v_x=1.09~\mbox{meV}$, $r=143~\mbox{nm}$, and $s=81.9~\mbox{nm}$, $\Theta$ is the Heaviside's step function, $v_y \gg v_x$, $\delta y_2$ is of a small value representing a very narrow wire, and $d=287~\mbox{nm}$ is the distance between the nano wire and the center of the coupled
quantum dots in the $y$ direction.    

{\em Analysis of the measurement:} Prior to the measurement, the
electron in the double dot and the electron in the nano-wire are spatially
well-separated, so the state of the combined system is $\rho \otimes
\ket{\psi}\bra{\psi}$, where $\rho$ is the state of the qubit and $\ket{\psi}$
is the state of the electron incident on the RT barriers. We take the state of
this electron to be a Gaussian wave-packet. We
include the first four eigenstates of the dot in our numerical simulation, but
for the sake of the following discussion, we will assume the dot contains only
the two computational states (which it {\em very} nearly does). After the
interaction, the state of the system is $U \rho \otimes \ket{\psi}\bra{\psi}
U^\dagger$, where the unitary operator $U$ acts in the
joint space, and the qubit and the electron become entangled. For a fictitious
observer who performs a von Neumann measurement of the momentum of the (now
outgoing) electron, the final state, on obtaining the momentum $p$, is a 
normalized version of
\begin{eqnarray} 
  \rho_p & = & \bra{p} \otimes I ( U \rho\otimes \ket{\psi}\bra{\psi} U^\dagger ) I \otimes \ket{p} \nonumber \\ 
         & = & A_p \rho A_p^\dagger 
\end{eqnarray} 
for some operators $A_p$ which satisfy $\int A_p^\dagger A_p dp = I$, and 
which completely characterize the measurement process. The probability that the
final momentum is $p$ is $\mbox{Tr}[A_p^\dagger A_p \rho]$, where this, and all subsequent traces, are taken over the qubit system. Of course the
measurement we actually make determines only whether the final momentum of the
qubit is positive or negative. The probability that we will detect a
transmission is thus $P_+ = \int_{0}^{\infty} \mbox{Tr} [A_p^\dagger A_p \rho]
dp$, that we will detect a reflection is $P_- = \int_{-\infty}^{0} \mbox{Tr}
[A_p^\dagger A_p \rho] dp = 1 - P_+$, and the corresponding final states are 
$\rho_+ = \int_{0}^{\infty} A_p \rho A_p^\dagger dp / P_+$ and $\rho_- =
\int_{-\infty}^{0} A_p \rho A_p^\dagger dp / P_-$. We note that the operators
$A_p$ are matrices with elements $A_p^{ij} = \bra{i}\bra{p} U \ket{\psi}\ket{j}$, 
(where the $\{\ket{i}\}$ are a basis for the electron in the double dot)
and can therefore be obtained using exclusively pure state simulations. 

From the numerical results we find that the interaction with the qubit causes
very  little change in the energy of the pumped electron, namely the scattering 
is essentially elastic. While we do not assume elastic scattering in
obtaining our results, such an assumption is conceptually useful. If the
scattering is completely elastic, then one can write the operators as $A_p =
\tilde{A_p}|\braket{p|\psi}|^2$ where the $\tilde{A_p}$ are independent of the
initial state. If we select the momentum of the incident electron so that it is
sharply peaked at $p$, then the only  operators which contribute to the
measurement are those at $p$ and $-p$. In this case the measurement is
described solely by the two operators $A_p$ and $A_{-p}$, so that the
probabilities are $P_+ = \mbox{Tr} [\tilde{A}_p^\dagger \tilde{A}_p \rho]$ and
$P_- = \mbox{Tr} [\tilde{A}_{-p}^\dagger \tilde{A}_{-p} \rho]$, and the final
states are $\rho_+ = \tilde{A}_p \rho \tilde{A}_p^\dagger dp/P_+$ and $\rho_- =
\tilde{A}_{-p}\rho \tilde{A}_{-p}^\dagger dp /P_-$.

If the location of the qubit in the double dot has a large effect on the
transmission coefficient of the barriers, then a single measurement cycle
(consisting of a pumping event followed by a measurement of the electrode) will
extract a lot of information about the state of the qubit in this basis.
However, in general both states give some chance of transmission and a single
cycle will not therefore discriminate completely between the basis states.
Nevertheless, if the measurement operators $A_p$ are diagonal in this basis,
then the measurement will cause no undesirable disturbance. In this case, a
complete von Neumann measurement could be approached arbitrarily closely by
simply repeating the pump/measurement cycle the required number of times.
However, the operators $A_p$ do not have this property, and therefore do cause
a disturbance which limits the total amount of information which can be
extracted simply by repetition alone. 

Using the polar decomposition theorem we can write the operator $A_p$ as the
product of a unitary $U_p$ and a positive operator $P_p = (A_p^\dagger
A_p)^{1/2}$, so that $A_p = U_pP_p$. It is $P_p$ which changes the entropy of the
system and thus performs the extraction of information. This information is
extracted in the eigenbasis  of $P_p$; that is, it is information about which
of the eigenstates of $P_p$ the system is in. The imperfection of the
measurement can therefore be understood as resulting because the positive
operator $P_p$ is diagonal in the wrong basis, and/or because $U_p$ causes an
additional disturbance. Now, if our measurement is described by only two
operators, then it is easy to show that the $P$'s for both are diagonal in the
same basis. As a result we can correct for any error in the measurement basis
by applying a unitary to select the correct basis. Secondly, upon obtaining
the measurement result, since we know which of the $U$'s has been applied
(being either $U_p$ or $U_{-p}$), we can correct for it by applying the
Hermitian conjugate after the measurement in what is a simple example of a
feedback control procedure~\cite{wiseman-fb,Belavkin,DJ,DHJMT,DHelon06}. Therefore, when the measurement has only
two operators we can correct completely for any undesirable disturbance, and 
obtain a complete von Neumann measurement by repetition. In our case such 
perfect correction is not possible because the scattering is not completely
elastic, and the momentum of the electron has a finite spread. Nevertheless, 
we find that we can improve the performance significantly by using an initial 
basis change and unitary feedback at each pump/measurement cycle designed to 
correct for the disturbance of the operators $A_p$ at the center of the
electron wave-function.

{\em Results:}
To quantify the accuracy of the measurement we proceed as follows. We encode
one  bit of information in the qubit using the ensemble consisting of the
computational basis states chosen with equal probabilities. We then calculate
the amount of information which the measurement fails to extract ${\mathcal F}
= 1 - M$, where $M$ is the mutual information in bits~\cite{Pierce}. This is
the residual uncertainty left after the measurement is made. The smaller
${\mathcal F}$ the more perfect the measurement.

\begin{figure}[t]
\centering
\includegraphics[width=3.3in]{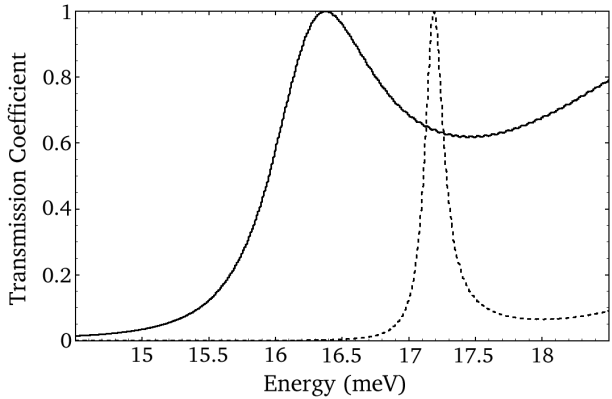}
\caption{The transmission coefficients of the RT barriers are plotted here as a function the 
energy of the pumped electron. The solid and dashed lines give the transmission when the 
qubit is in the states $\ket{0}$ and $\ket{1}$, respectively.}
\label{trans}
\end{figure}

\begin{figure}[t]
\centering
\includegraphics[width=3.3in]{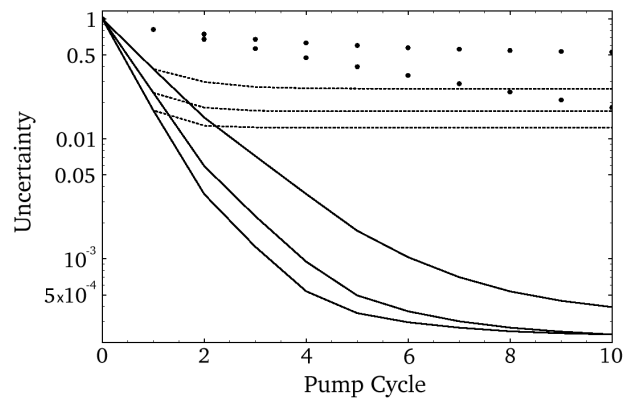}
\caption{The residual uncertainty is plotted here for a single-qubit
measurement using a sequence of pump/measurement cycles. Each curve shows the
increase in accuracy as the number of pump cycles is increased. The dashed
lines are measurements without feedback, and the solid lines with feedback, both 
using an incident electron energy of $E=16.4~\mbox{meV}$. The
three curves for each case, from bottom to top, are for incident energy uncertainties   
of $\Delta E = 2\%$, $2.8\%$ and $4.2\%$. For comparison, the solid circles represent the residual uncertainty for
$E=17.6~\mbox{meV}$ with $\Delta E = 2\%$.}
\label{MI}
\end{figure}

We now examine the transmission profile of the RT barriers as a function of the
incident energy for both the states $\ket{0}$ and $\ket{1}$, as shown in
Figure~\ref{trans}. We find that at energy $E=16.4~\mbox{meV}$, the RT barriers provide 
a highly sensitive meter of the qubit state. The residual uncertainty, $F$, for 
a measurement using a pumped
electron with this average energy is plotted in Figure~\ref{MI}.  Up to 10 
repetitions of the pump
cycle, both with and without the correcting feedback, are employed.  We also 
calculate this residual uncertainty for a range of values of the energy uncertainty, 
$\Delta E= 2\%, 2.8\%$ and $4.2\%$ of the mean energy. Current experiments with electron pumps indicate that spreads at 
least this narrow should be 
achievable~\cite{Likharev2001, jehl2003}.  The dashed lines give the results without feedback (for six
values of the energy uncertainty), and we see in this case that the first pump
cycle extracts up to about $90\%$ of the information, and subsequent cycles extract 
virtually no
further information, due to the disturbance caused by the first cycle. However,
with the simple feedback procedure described above, which includes an initial
rotation, the second pump cycle extracts considerably more information, as do
subsequent pump cycles, plateauing at about the 6th cycle. The result is a
measurement with high accuracy, in which the residual uncertainty is less than
two parts in a thousand. For comparison we also evaluate the measurement obtained
when the incident energy is $E=17.6~\mbox{meV}$ with an energy spread $\Delta E=2\%$. 
In this case the transmission is not as sensitive to the qubit state, and $10$ 
measurement cycles are not sufficient to reduce the residual uncertainty bellow $15\%$.   

{\em Acknowledgments:} We acknowledge the use of the supercomputing resources 
at the Queensland Parallel Supercomputing Facility and iVEC The hub of advanced computing in WA. This work was supported by 
the ARC, The University of Western Australia and the State of Queensland.

\end{document}